# Subjects classification from high-dimensional and small-sample size datasets using a strategy based on "Clustering Variables around Latent Components (CLV)" method.

Dimitri Marques Abramov

Laboratory of Neurobiology and Clinical Neurophysiology, National Institute of Women, Children and Adolescents Health Fernandes Figueira, Oswaldo Cruz Foundation, Rio de Janeiro, Brazil

dimitri.abramov@iff.fiocruz.br

**Abstract**

*High-dimensional complex systems can be studied through multivariate analysis, as Principal Component Analysis, however large samples of observations frequently are needed for it. Here it is examined a method for small samples based on clustering variables around latent variables (CLV) to subject classification in two presumed groups. For it, a predictive model was developed to generate datasets with two groups of cases whose variables show randomness features (up to 30% of variables manifest difference between groups, and up to 7% of those are correlated between them). The method recovered the information of the latent factors to classify the subjects with 80 to 95% of agreement, with positive relationship between the classifier precision and the rate [number of variables / number of subjects].*

# 1. Introduction

The multivariate analysis (MVA) is a powerful statistical strategy for the analysis of large datasets, with variables in complex interrelationships that must be studied as a whole (1, 2). With MVA, patterns and dynamics of a set of cases can be mapped by observing many variables (1). One of the objectives is effective clustering for a specific classification of observed cases (3). Two major fields of application of MVA are social sciences and psychologies, but also exploratory genetics and neurosciences. MVA allows the study of complex systems in general, but for this it needs many dimensions of observation (variables) .

The MVA assumes that a multidimensional system has variables that are not directly observable and measured but that determine the behavior of the system as a whole. These variables are latent components (or factors), which are incorporated into the variables that were collected. The interdependence between variables is related to the presence of information of each of these latent factors, to a greater or lesser extent, in each variable collected (2, 4). This multidimensional dataset is generically determinable by the function (4):

$$X_{i,s} = 1/J \sum(F_j * L_{i,j}) + \varepsilon_{i,s} \quad [1]$$

Where X is a matrix of observations of the measurable I variables ($i$ = 1, ..., I) for the cases in a sample S ($s$ = 1, ..., S), which is related to the integration of the hidden J factors in F ($j$ = 1, ..., J, where J < I) multiplied by the correlation weights in L that determine how much of each factor $j$ is embedded in each variable $i$. Sum up this integration to a sample a random individual error, specific for each observation $s$ of each observed variable $i$, this error whose mean tends to zero and variance to 1.

The major objective of the MVA is to somehow access this axial set of factors in F. For this, conventional methods such as Principal Component Analysis require a considerable number of observations (5,6). On the other hand, the greater the influence of the latent factors on each variable, the less expressive becomes the effect of the size of the sample (5). That is, datasets whose variables do not show robust interdependence should contain considerably more cases than variables.

Vigneau and Qannari (7, 8) developed a hidden factor analysis method (they call "clustering variables around latent variables" - CLV) for datasets with more variables than observations, based on the hierarchical clustering of the variables through a similarity approximation method, that is, a non-Euclidean distance measure based on the correlation between two variables (8). Latent factors are extracted from the mean of the standardized variables, within each group at a level of the hierarchical tree (7).

Here, we have developed a model to verify the ability of the Vigneau and Qannari's method to be used to reduce the dimensionality of a dataset to its latent factors, allowing the classification of cases in a predetermined number of groups, through the method of partitioning K-means.

The objective of this study is to provide a possible tool for validation of diagnoses related to complex conditions, such as psychiatric disorders, through a large dataset of objectively known variables (such as biological and psychometric variables), aiming to classify the subjects of a small sample as belonging to the groups "cases" and "control", allowing the evaluation of validity of the clinical diagnosis.

In practice, the use of conventional factor analysis methods (such as factor analysis) requires a very large number of subjects, in proportion to the size of the dataset. For example, following the 10: 1 rule, 300 variables would require at least 3000 subjects (9), equivalent to 900,000 biochemical or functional tests, for an adequate determination of latent factors and consequent dimensional reduction. This is usually a practical impossibility, even for conditions of high prevalence.

## 2. Methodology

2.1. Mathematical model for datasets

From the function [1] is implemented the equation:

$$x_{i,s} = [\ 1/J \sum(k * f_{s,g,j} * l_{i,j})] + \mu_i + (\mu_i * \varepsilon_{i,s}) \qquad [2]$$

Where I = 50, I = 100 or I = 300; S = 40, divided into two groups G, such that G1 = {s = 1, ... 20} G2 = {s = 21, ..., 40}. We adopted J = 4, J = 6 and J = 8 latent factors in F, where each factor f is a vector with 20 random positive and negative values of normal distribution, such that:

$$\mathbf{f}_{2j} = \mathbf{f}_{1j} + [\max(\mathbf{f}_{1j}) - \min(\mathbf{f}_{1j})] * 1.1 \qquad [3]$$

which establishes two sets of factors that unambiguously determine the subjects of groups G1 and G2 through the k-means method. The eight factors used were ordered and plotted in Figure 1.

. The parameter *k* determines the relative power of the latent factors, where k = 0, 0.1, 0.2, ...,1. The weigths $l$, for each *i* and *j*, determine the proportional load of each latent factor in each variable in X. The weights in L are determined by:

$$L = (M * q) + (1-q) \qquad [4]$$

Where M is an I x J matrix of random numbers between 0 and 1 and q = 0.25, making the values in L vary between 0.75 and 1. This strategy ensures that all variables in X have at least 75% of each factor incorporated into them, which limits the heterogeneity between the variables.

We add considerable noise through a mean $\mu$ for each variable i, which is:

$$\mu_i = m_i^2 \qquad [5]$$

Where m randomly varies between 1 and 10 in a normal distribution, plus to na error $\mu*\varepsilon$, where $\varepsilon$ also randomly varies between -2 and 2, being mean that tends to zero.

It is generated 50 datasets for each value of I, J and k.

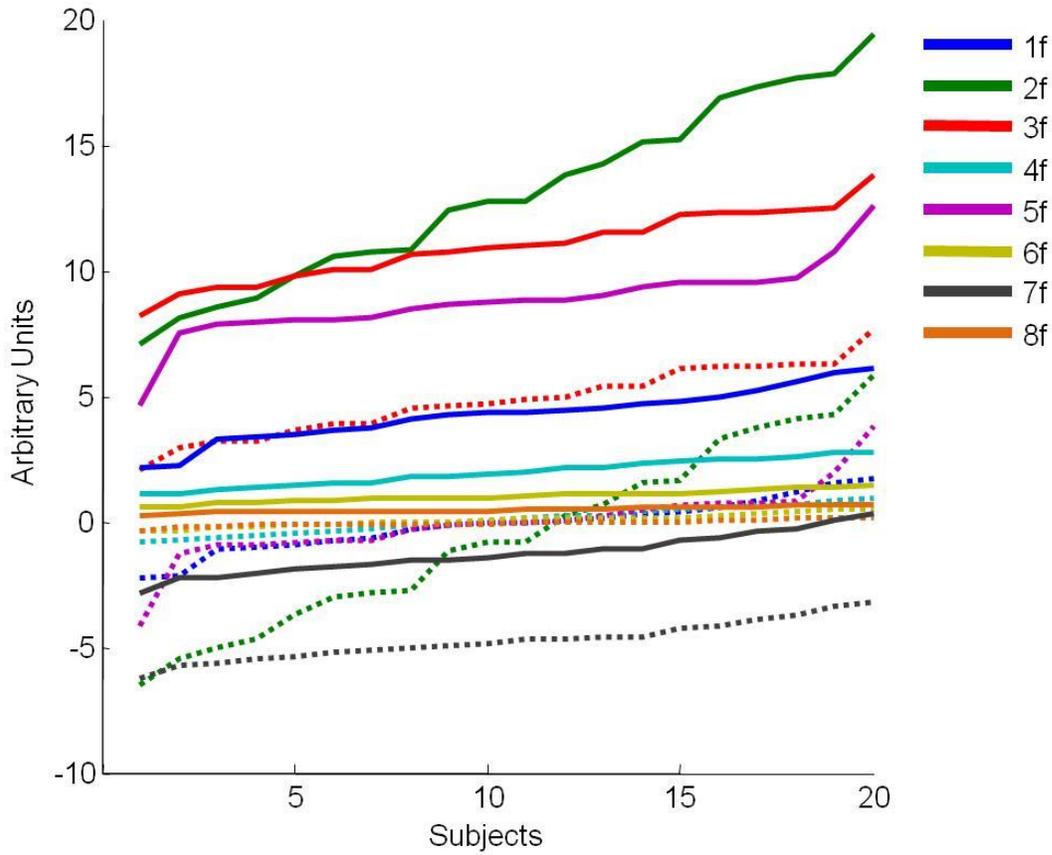

Figure 01. The eight latent factors used. Factors in dotted lines belong to group 1 and in continuous lines to group 2.

## 2.2. CLV classification methodology.

The dataset variables were clustered by the hierarchical method that uses the Pearson's correlation coefficients as distance measure, according to (10):

$$d_{a,b} = 1 - \frac{(\mathbf{x}_a - \mu_a)(\mathbf{x}_b - \mu_b)'}{[(\mathbf{x}_a - \mu_a)(\mathbf{x}_a - \mu_a)']^{\frac{1}{2}} [(\mathbf{x}_b - \mu_b)(\mathbf{x}_b - \mu_b)']^{\frac{1}{2}}}$$

[6]

given that $\mathbf{x}_a$ and $\mathbf{x}_b$ are two vectors (columns) in the dataset X, which correspond to two variables with means $\mu_a$ and $\mu_b$, where $a = 1, ..., I$, $b = 1, ..., I$. If two variables are negatively correlated, $d_{a,b}$ tends to 2, while if they are positively correlated, $d_{a,b}$ tends to 0.

Since all distances are between 0 and 2 (are dimensionless indices relative to the Pearson's coefficients), a geometric strategy is admissible as a linkage process determining the centroids of the grouped distances. We adopted Ward's method (11), according to:

$$d'_{A,b} = \left[ \frac{2n_A n_b}{n_A + n_b} (d_{a,b}^2 + d_{a',b}^2 - d_{a',a}^2) \right]^{1/2} \quad [7]$$

Where $d'$ contains the new distances for the new cluster $A$ and the other clusters $b$, determined by the original distances $d_{a,b}$ and $d_{a',b}$ of the previous hierarchical level incremented by the increase in cluster size, where $n$ is the number of elements in those clusters $b$ and the new cluster $A$, $a$ and $a'$ are the clusters that form the new cluster $A$.

Similar to the original work of Vigneau and Qannari (7), we extracted a new variable from each cluster, which we call here "Resultant Vectors" (RV). We compute the RV sets of the variables from the first to the fifth level of the hierarchical tree, providing, respectively, two, three, four, five and six RVs. Each RV is determined by the equation:

$$\mathbf{rv} = 1/N \sum (\mathbf{x}_i - \mu_i) / \sigma_i \quad [8]$$

Where $\mathbf{x}_i$ is the vector of values of the variable with all cases, where N is the number of all the variables $i$ of the respective cluster, $\mu$ and $\sigma$ are the mean and standard deviation of the vector $\mathbf{x}_p$.

The sets of two to six RVs are inputs to the k-means algorithm, defining two clusters for classification.

## 2.3. Statistical analysis

Each dataset results in a vector with subjects reclassification indexes in one of the two groups obtained through the k-means method, which is compared with the original indexes, generating a congruence coefficient ranging from 20 (50%, random reclassification, what is expected for k = 0) to 40 (100%, perfect match, non-randomic).

The datasets for each set of parameters determine groups with 50 congruence coefficients, which will be compared statistically using one way analysis of variance (ANOVA-1).

We explored the datasets generated with six latent factors and 300 variables regarding the behavior of groups and variables as a function of each value of $k$. We found in each dataset the value of the probability of equality between the two groups for each variable through the Mann-Whitney U Test. We surveyed how many comparisons obtained $p < 0.05$, in each dataset. Also, we sequentially correlated each of the first 150 variables with each of the last 150 variables by the Pearson's method, calculating how many correlations were significant ($p < 0.01$) and the mean value of the correlation coefficients, for each dataset.

## 3. Results

All groups for the parameters k, number of latent factors J and number of variables I resulted in similar relationships as a function of the number of RVs used: six RVs produce the best congruence results in comparison to the other RV sets, as we see in figure 2, for 6 latent factors, as $k$ ranges from 0 to 1. In this case, the averaged congruency ranged from 32 (80%, for k = 0.1) to 38 (95%, for k = 1) subjects, using 6 RV. Both the RV number and the $k$ value result in statistically different congruences (respectively, $F = 6.28$, $p < 0.0001$, and $F = 23.19$, $p < 0.0001$).

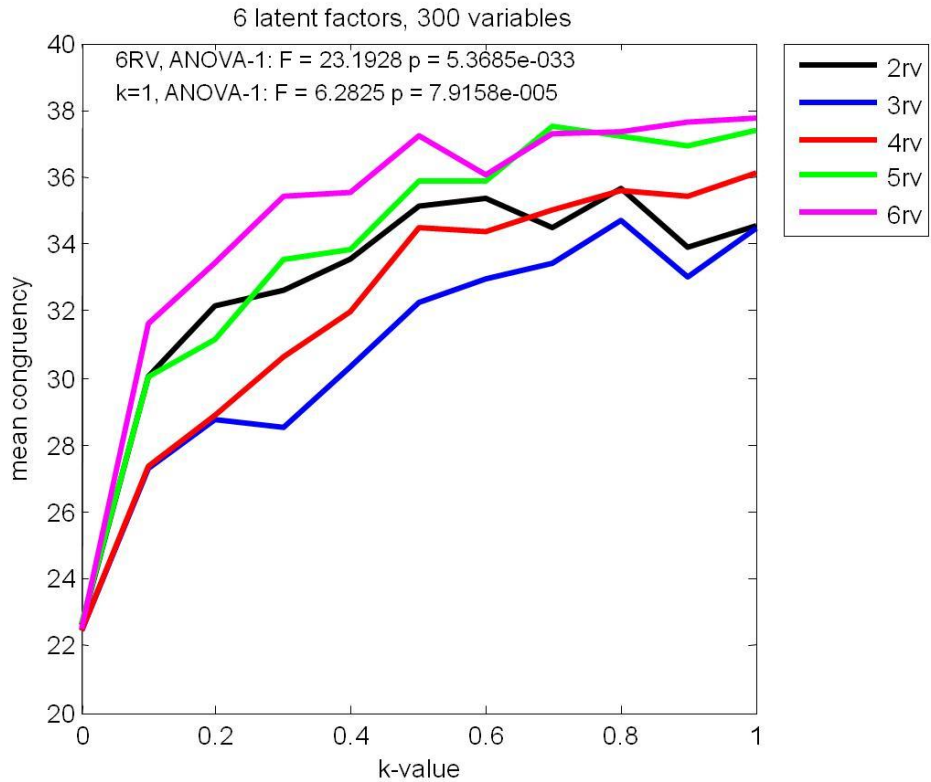

Figure 2. Averaged congruency for 6 latent vectors and 300 variables. ANOVA-1 comparisons between results for k values from 0.1 to 1 (by 6RV, above) and among the RV groups (by k = 1, bellow).

The number of latent factors (figure 3) does not seem to influence the averaged congruence, independent of the number of RVs found (for k = 1 and 300 variables in the dataset). However, with 2RV, the number of latent factors tends to influence negatively the averaged congruency.

However, the number of variables in the dataset positively influences the mean congruences (figure 4, for k = 1 and 6 latent factors), regardless of the number of RVs used.

The descriptive statistics of the datasets (six latent factors, 300 variables, and *k* values from 0 to 1) show that on average, 12 to 30% of the variables showed significant differences between the groups for each dataset generated by the model. In the 150 comparisons between variables by dataset, from 1.5 to 7% of these correlations revealed statistical significance (p < 0.01, Pearson's test), and the mean correlation coefficients ranged from 0.13 to 0.17.

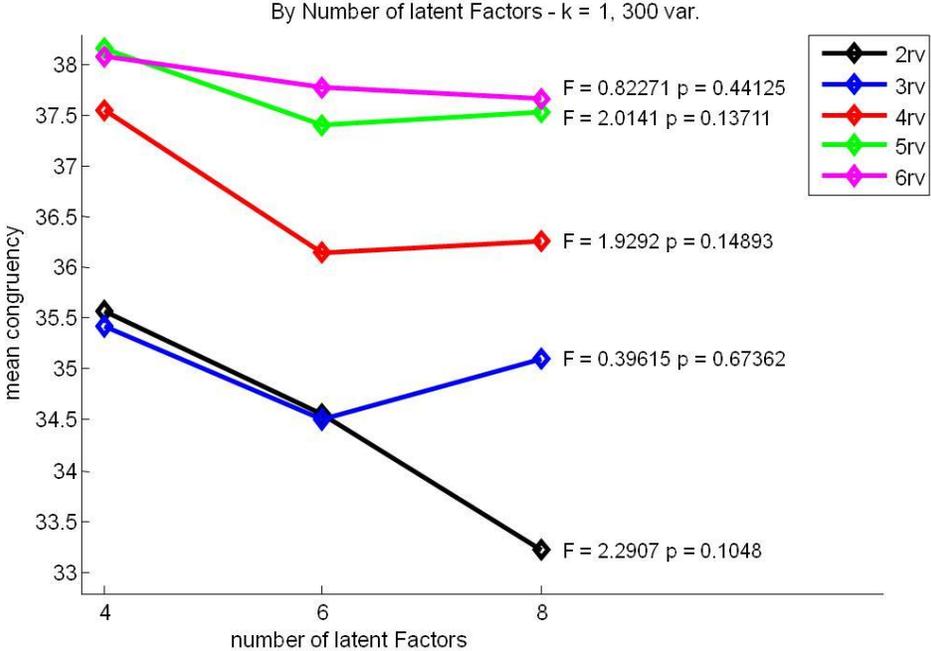

Figure 3. averaged congruency by number of latent factors, for each RV set, k = 1 and 300 variables. Respective ANOVA-1 F and p-values close of each dataplot.

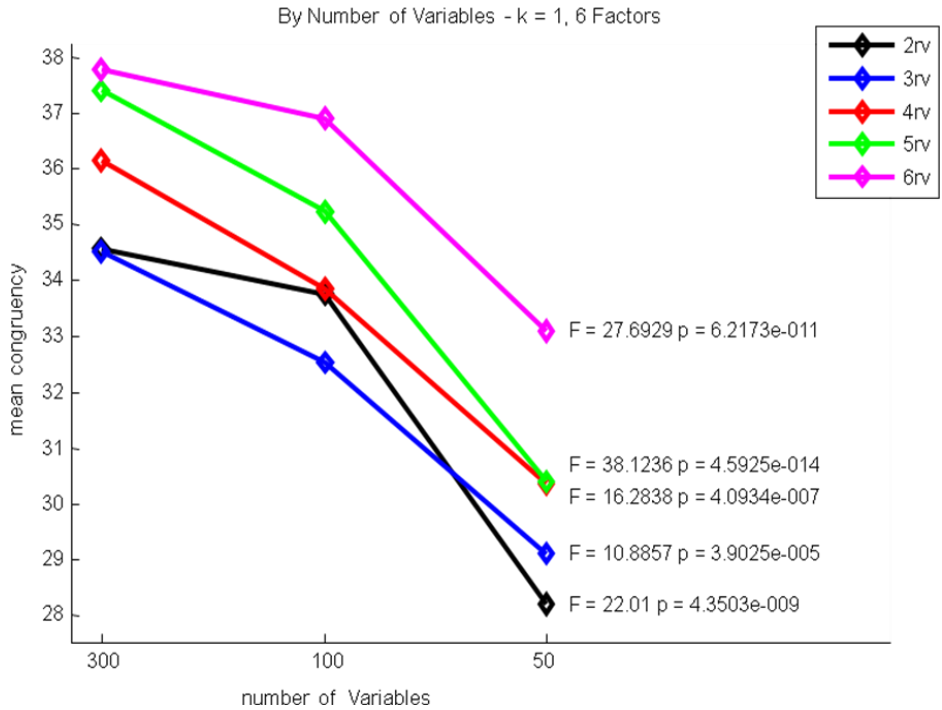

Figure 4. Averaged congruency in function of the number of variables in the dataset, for k = 1 and 6 latent factors. Respective ANOVA-1 F and p-values close of each respective dataplot.

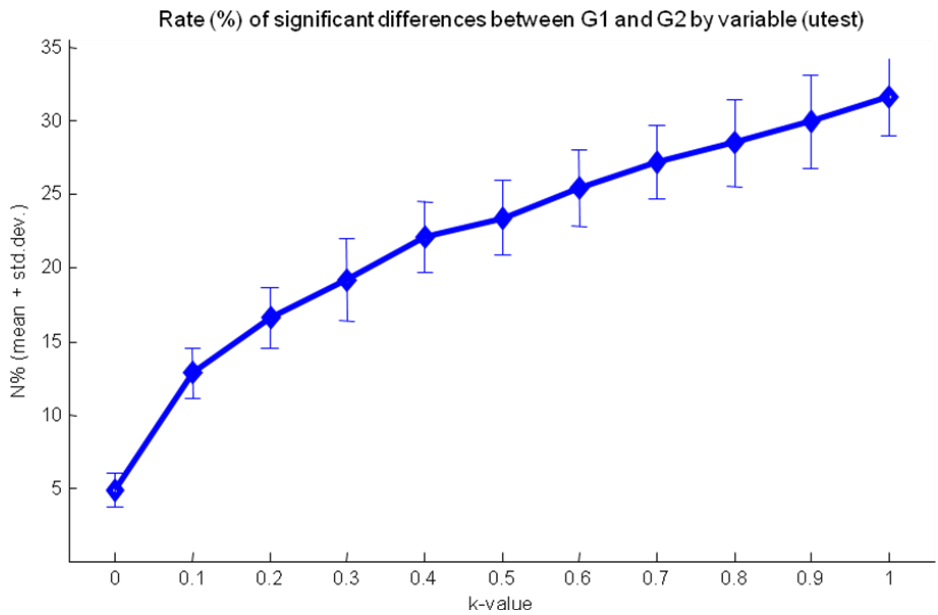

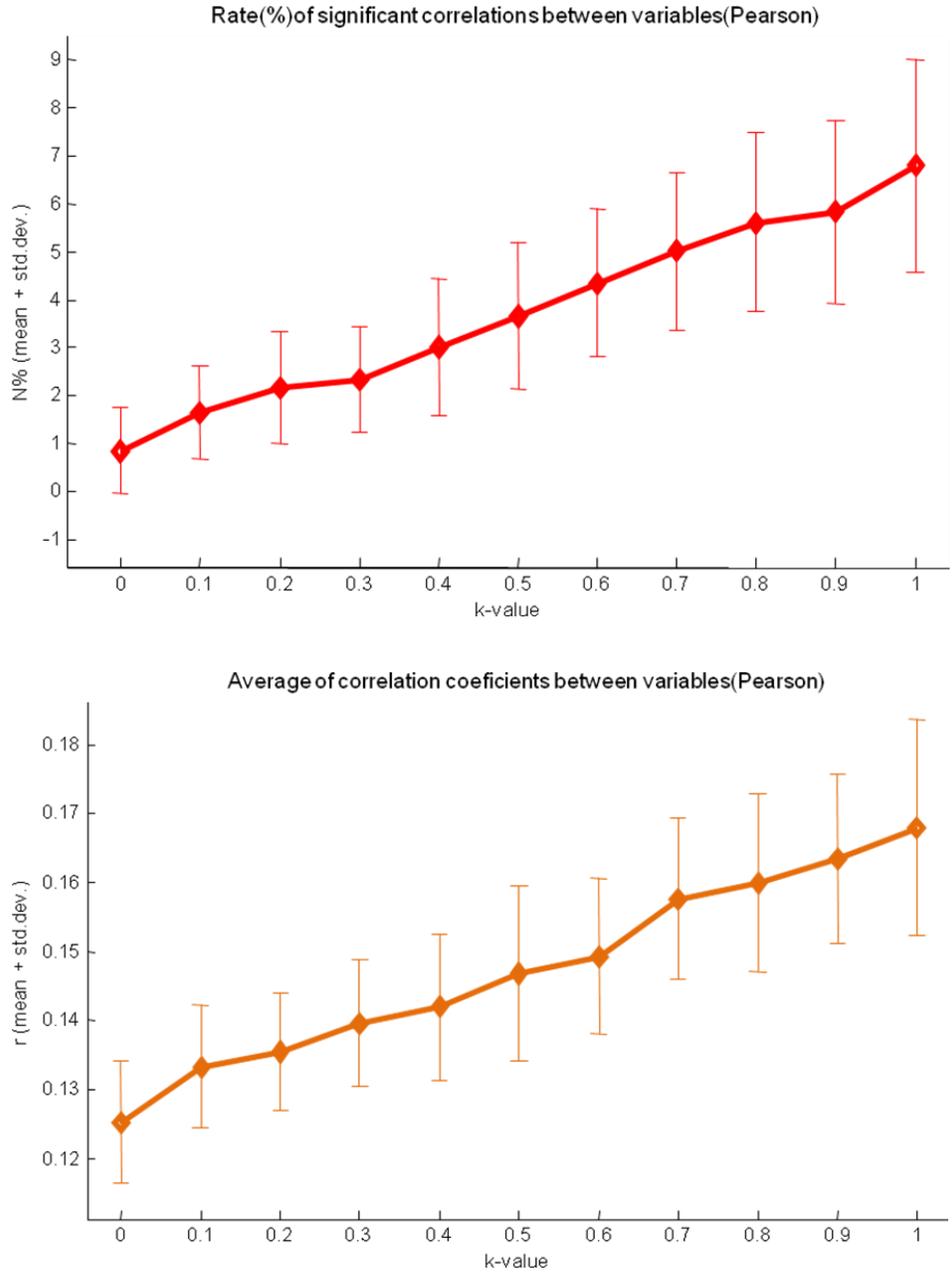

Figure 5. Descriptive statistics for the datasets, by each *k* value, for 6 latent factors and 300 variables. Upper panel (blue), the averaged rate (%) of statistical differences between groups by variable (n = 300, p <0.05, by U-test). Middle panel (red), averaged rate (%) of significant correlation coefficients (n = 150, p <0.01, by Pearson Test, 150 comparisons by dataset). Bottom panel (orange), averaged Pearson's correlation coefficients (n = 150). Ploting means and standard deviations.

## 4. Discussion

From the dataset generated by the model, the presented classification method, based on the CLV strategy, recovers significant amount of latent information for reclassify the subjects, according to the parameters used. The results are optimal for 300 variables and k = 1 when the classifier shows a 95% efficiency using six RVs (at the fifth level of hierarchical clustering) regardless of the amount of latent factors used.

The model used arbitrary parameters, such as *k*, the range of latent factors, sample means and errors, as well as the distribution of weights for composition of variables. The objective was to generate datasets of apparent randomness, which show few significant differences between the groups in relation to their variables, and specially low correlation between variables. Under these conditions, the method was appropriate for extracting information and effectively conducting reclassification. There is sufficient evidence for the method to be considered. The ability of the test to reclassify subjects will be proven in the real world, in case it reclassifies with great accuracy subjects previously diagnosed by a gold standard method.

Although the model is arbitrary for its various parameters, it shows that the efficacy of the classification method is proportional to the total number of variables, or to the ratio [number of variables / number of cases], unlike other factor analysis strategies. This finding opens the door to a new tool for analysis of high dimensional, complex systems using a relatively small set of observations, a tool that deserves to be further studied.